\documentclass[aps,prb,10pt,preprint,groupedaddress,nofootinbib,showkeys]{revtex4-1}

\usepackage[dvips]{graphicx}
\usepackage[english]{babel}
\usepackage[usenames]{color}
\usepackage[normalem]{ulem}
\usepackage{amsfonts}
\usepackage{amssymb}
\usepackage{amsmath}
\usepackage{textcomp}
\usepackage[up]{subfigure}
\usepackage[squaren,thickqspace,mediumspace]{SIunits}
\usepackage[colorlinks=true,linkcolor=blue]{hyperref}
\usepackage{url}
\usepackage{dcolumn}

\begin{document}

\title{Atomistic spin dynamics of low-dimensional magnets}

\author{Lars Bergqvist}
\email{lbergqv@kth.se}
\affiliation{Department of Materials Science and Engineering and Swedish e-Science Research Center (SeRC), KTH Royal Institute of Technology, Brinellv. 23, SE-100~44 Stockholm, Sweden} 
\author{Andrea Taroni}
\affiliation{Department of Physics and Astronomy, Uppsala University, P.O. Box 516, 751~20 Uppsala, Sweden}
\author{Anders Bergman}
\affiliation{Department of Physics and Astronomy, Uppsala University, P.O. Box 516, 751~20 Uppsala, Sweden}
\author{Corina Etz}
\affiliation{Department of Physics and Astronomy, Uppsala University, P.O. Box 516, 751~20 Uppsala, Sweden}
\author{Olle Eriksson}
\affiliation{Department of Physics and Astronomy, Uppsala University, P.O. Box 516, 751~20 Uppsala, Sweden}

\date{\today}

\begin{abstract}
We investigate the magnetic properties of a range of low-dimensional ferromagnets using a combination of first-principles calculations and atomistic spin dynamics simulations. This approach allows us to evaluate the ground state and finite temperature properties of experimentally well characterized systems such as Co/Cu(111), Co/Cu(001), Fe/Cu(001) and Fe/W(110), for different thicknesses of the magnetic layer. We compare our calculated spin wave spectra with experimental data available in the literature, and find a good quantitative agreement. We also predict magnon spectra for systems for which no experimental data exist at the moment, and estimate the role of temperature effects.
\end{abstract}

\keywords{Atomistic spin dynamics; monolayer magnetism; ultrathin magnets; magnons; spin wave; iron; tungsten; cobalt; copper}

\maketitle

In recent years, spin-polarized electron energy loss spectroscopy (SPEELS) has been developed to the point of becoming a powerful method for investigating spin waves at surfaces and in thin films.~\cite{Plihal1999} Landmark experiments have included the measurement of the magnon spectrum of ultrathin Co films on Cu(001),~\cite{Vollmer2003} and of a single monolayer (ML) of Fe on W(110).~\cite{Prokop2009} More recently, an asymmetry in the magnon spectrum of 2 ML Fe on W(110) has also been reported~\cite{Zakeri2010} - this is a direct signature of the Dzyaloshinskii-Moriya interaction present in this system.~\cite{Udvardi2009} The experimental accessibility of these properties provides an opportunity for comparing the merits of different theoretical approaches used to describe thin-film magnets.~\cite{Udvardi2003,Costa2008,Bergman2010} Furthermore, the experiments challenge theoreticians to address issues such as accurately characterizing exotic ground states arising from relativistic spin-orbit coupling effects,~\cite{Udvardi2009,Bode2007,Heide2008} and treating finite-temperature effects, for example.

Previous theoretical studies regarding the spin-wave dispersion in thin layers have been performed by Costa et al.~\cite{Costa2006,Costa2008}. The method used to investigate these systems consists of a tight binding description, either from empirical data or from first principles calculations, of both the magnetic layers and the substrate followed by a dynamical theory of the spin-waves.~\cite{Costa2004a,Costa2004b,Costa2010,Mills2012}
Recently the formalism of atomistic spin-dynamics (ASD) was developed~\cite{Antropov1996} and was subsequently implemented in the UppASD package~\cite{Skubic2008,uppasd}. With this technique it is possible to model spin waves at finite temperature, as shown recently for 1 ML of Fe on W(110).~\cite{Bergman2010} The method relies on an accurate estimation of the exchange parameters in the Hamiltonian using first-principles calculations, and subsequently evolving it in time according to a modified Landau-Lifshitz-Gilbert (LLG) equation of motion,~\cite{Skubic2008,Mentink2010} as described in the next section. 

In this article, we extend the scope of our previous investigations, and employ the first-principles/ASD strategy to simulate the magnon spectra of Co/Cu(001), Co/Cu(111),  Fe/Cu(001) and Fe/W(110), for varying thicknesses of the magnetic layer. The chosen systems have been extensively studied experimentally and recently, using SPEELS, also magnon spectra have been measured for selected systems with certain thicknesses of the magnetic overlayer.

\section{Theoretical Approach}
\subsection*{First Principles Calculations and Atomistic Spin Dynamics Simulations}

Our strategy is to combine first-principles calculations with ASD simulations. In practice, we map an itinerant electron system onto an effective Hamiltonian with classical spins,
 \begin{equation*}
 \mathcal{H} = \mathcal{H}_{\mathrm{ex}} + \mathcal{H}_{\mathrm{SO}}. 
 \end{equation*}
 This generalized Hamiltonian consists of an exchange term:

\begin{equation}
  \mathcal{H}_{\mathrm{ex}} = -\sum_{i\neq j} J_{ij} \mathbf{m}_i \cdot \mathbf{m}_j,
\label{eqn:hh}
\end{equation}

\noindent where $i$ and $j$ are atomic indices, $\mathbf{m}_i$ is the classical atomic moment and $J_{ij}$ is the strength of the exchange interaction, and a spin-orbit interaction term:

\begin{equation}
 \mathcal{H}_{\mathrm{SO}} = K \sum_{i}\left(\mathbf{m}_i \cdot \mathbf{e}_K\right) ^{2} - \sum_{i\neq j} \mathbf{D}_{ij} \left(\mathbf{m}_i \times \mathbf{m}_j\right),
\label{eqn:so}
\end{equation}

\noindent where $K$ is the strength of the anisotropy field along the direction $\mathbf{e}_K$ and $\mathbf{D}_{ij}$ is the Dzyaloshiskii-Moriya (DM) vector.~\cite{Dzyaloshinskii1957,Moriya1960} These represent the main relativistic effects present in low-dimensional magnets, although the magnitude of the DM term is often negligibly small.

The ASD simulations rely on an accurate determination of the exchange integrals and spin-orbit coupling present in the generalized Hamiltonian. 
In order to determine these, we employed two different methodes for our first principles calculations. First, we set up a slab with 9-11 layers representing the substrate, having on one side the magnetic over-layer and a large region of vacuum, simulating the surface. The interlayer distances were relaxed according to the forces using the projector augmented wave method (PAW) as implemented in the VASP program~\cite{Kresse1996,Kresse1999}. Once the relaxed geometry is established, we employed the SPR-KKR package~\cite{EKM11,SPR-KKR5.4} for calculations of the parameters to be used within the generalized Hamiltonian using the local spin density approximation (LSDA) for the exchange-correlation potential. The basis set consist of $s$, $p$, $d$ and $f$ orbitals and all relativistic effects are included by solving the Dirac equation. For the selfconsistent calculation we employed between 4-500 k-points in the two dimensional Brilloiun zone, while a much more dense k-point grid of 2000 k-points were used in the calculation of exchange parameters and Dzyaloshinskii-Moriya interactions, sufficient to obtain a convergence on a micro Rydberg level for these parameters. 

The exchange parameters $J_{ij}$ and Dzyaloshinskii-Moriya (DM) vectors $\mathbf{D}_{ij}$ are obtained from the relativistic generalization~\cite{Udvardi2003,Ebert2009} of the real-space method of infinitesimal rotations of Liechtenstein, Katsnelson and Gubanov (LKG) ~\cite{Liechtenstein1984,Liechtenstein1986} with the ferromagnetic configuration chosen as the reference state for mapping. Here, the full exchange tensor $\bar{J}_{ij}$ is calculated which can be decomposed into its isotropic part $J_{ij}$ and its anti-symmetric part, which correponds to the DM-vector. The whole procedure relies on an adiabatic approximation in which the slow motion of the spins is decoupled from the fast motion of the itinerant electrons, a situation that is justified at low energy scales and for systems with reasonably large exchange splitting.

Once calculated, the exchange parameters $J_{ij}$ may be Fourier transformed to obtain the so-called adiabatic magnon spectrum. In the simple case of a single layer (corresponding to one atom per cell), the energy of a spin wave with respect to a ferromagnetic ground state is given by 

\begin{equation}
  E(\mathbf{q}) = \sum_{j\neq 0} J_{0j}\left[ \exp\left( i \mathbf{q}\cdot\mathbf{R}_{0j} \right) -1 \right],
\end{equation}
 
\noindent where $\mathbf{R}_{ij}$ is the relative position vector connecting sites $i$ and $j$. From this it is straightforward to calculate the spin wave dispersion $\omega(\mathbf{q})$.~\cite{Kubler2009} For systems with more than one atom per cell, as is the case for thin films consisting of more than one monolayer, the spin wave energies are given by the eigenvalues of the general $N \times N$ matrix here expressed in block form
%
\begin{equation}
  \begin{bmatrix} 
     \sum_j^N J_0^{ij} - J^{ii}(\mathbf{q}) & -J^{ij}(\mathbf{q}) \\
     -J^{ij}(\mathbf{q})^{\ast} & \sum_i^N J_0^{ji} - J^{jj}(\mathbf{q}) 
  \end{bmatrix},
  \label{eqn:Jijmatrix}
\end{equation} 

\noindent where $N$ is the number of atoms per cell (\emph{i.e.} in this case the number of monolayers). 


 Using in the ASD simulations the generalized Hamiltonian $\mathcal{H}$ as a starting point, the effective interaction field experienced by each atomic moment $\mathbf{m}_i$ is given by
\[
  \mathbf{B}_i = - \frac{\partial\mathcal{H}}{\partial\mathbf{m}_i}.
\]

\noindent The temporal evolution of the atomic spins at finite temperature is governed by Langevin dynamics, through coupled stochastic differential equations, the Landau-Lifshitz-Gilbert (LLG) equations, written here in the Landau-Lifshitz form, 

\begin{equation}
\frac{\partial\mathbf{m}_i}{\partial t} = - \frac{\gamma}{(1+\alpha ^2)}  \mathbf{m}_i \times \left[\mathbf{B}_i + \mathbf{b}_i(t)\right] - \gamma\frac{\alpha}{m (1+\alpha ^2)} \mathbf{m}_i \times \left\{ \mathbf{m}_i \times\left[\mathbf{B}_i + \mathbf{b}_i(t)\right] \right\},
\label{eqn:LLG}
\end{equation}

\noindent where $\gamma$ is the electron gyromagnetic ratio. Temperature fluctuations are included via a stochastic Gaussian shaped magnetic field $\mathbf{b}_i(t)$ with properties $\langle \mathbf{b}_i(t) \rangle = 0 $ and  

\begin{eqnarray} 
 \langle b_i^k (t) b_j^l (ẗ́') \rangle  = 2 D \delta _{ij} \delta _{kl} \delta (t-t'), & ~~
D=\dfrac{\alpha}{(1+\alpha ^2)} \dfrac{k_{\rm B} T}{\mu_B m},
\end{eqnarray}
where $i$ and $j$ denote lattice sites, $k$ and $l$ the carteisian components and $\alpha$ is the Gilbert damping parameter which eventually brings the system to thermal equilibrium. It should be noted that the simulations carried out in this work are for atomistic spins. Hence the gyromagnetic factor in this simulations is simply the ratio between the magnetic moment and the angular momentum of an atom. An anisotropy in the gyromagnetic factor, with respect to the orientation of the atomic spin, would appear in the first principles part of our calculation, where the spin and orbital magnetic moments are calculated. It is known that this effect is rather small for transition metals (e.g. as reported by Hjortstam \emph{et al}.~\cite{Hjortstam1997} and St\"ohr~\cite{Stohr1999}). The dependence on any possible anisotropy of the damping parameter is less known. However, for most of our calculations the thermal fluctuations do not force the moments to deviate too much from the easy magnetization axis, and hence a possible tensorial form of the damping parameter would not influence our results by a significant amount.

The coupled equations of motion (\ref{eqn:LLG}) can be viewed as describing the precession of each spin about an effective interaction field, with complexity arising from the fact that, since all spins are moving, the effective field is not static. In our calculations we evolve the stochastic LLG equations using a semi-implicit method introduced by Mentink \emph{et al}.~\cite{Mentink2010}

The principal advantage of combining first-principles calculations with the ASD approach is that it allows us to address the dynamical properties of spin systems at finite temperatures.~\cite{Skubic2008,Tao2005,Chen1994} We focus in particular on two important quantities, the space- and time-displaced correlation function: 

\begin{equation}
  C^k (\mathbf{r}-\mathbf{r'},t) = \langle m_{\mathbf{r}}^k(t) m_{\mathbf{r'}}^k(0) \rangle - \langle m_{\mathbf{r}}^k(t) \rangle \langle m_{\mathbf{r'}}^k(0) \rangle,
\end{equation}

\noindent where the angular brackets signify an ensemble average and $k$ the cartesian component, and its Fourier Transform, the dynamical structure factor:

\begin{equation}\label{structure_factor}
  S^k(\mathbf{q},\omega) = \frac{1}{\sqrt{2\pi}N} \sum_{\mathbf{r},\mathbf{r'}} e^{i\mathbf{q}\cdot(\mathbf{r}-\mathbf{r'})} \int_{-\infty}^{\infty} e^{i\omega t} C^k (\mathbf{r}-\mathbf{r'},t) dt,
\end{equation}

\noindent where $\mathbf{q}$ and $\omega$ are the momentum and energy transfer, respectively. The dynamical structure factor, $S(\mathbf{q},\omega)$ is the quantity probed in neutron scattering experiments of bulk systems~\cite{Lovesey1984}, and can analogously be applied to SPEELS measurements. By plotting the peak positions of the structure factor along particular directions in reciprocal space, the spin wave dispersions may be obtained.~\cite{Bergman2010,Skubic2008,Tao2005,Chen1994} 
%

\section{Results}
\subsection{\textit{ab initio} calculated quantities}
\label{abinitio}
As stated in the previous paragraphs, we use in our study of magnetic layers a combined first principles/ASD strategy. This means that we first lead an \textit{ab initio} investigation of the electronic structure and magnetic properties of all the systems under consideration. In the following, we shall state for all the studied systems the layer-resolved magnetic moments and the values of the exchange parameters. The exchange interactions have been calculated for a distance equal to three times the lattice parameter for each system. In Tables~\ref{tableFe} and \ref{tableCo} only the highest values of the exchange parameters are presented, namely the inter- and intra-layer nearest neighbours interactions. 
%

\begin{table}[h]
\caption{Exchange parameters and magnetic moments as calculated from first principles, for the Fe multilayers systems. All the exchange interactions up to a distance of three lattice parameters have been calculated and included in simulations. For clarity, we list only the dominant inter- and intra-layer interactions in the table (all the other exchange paramters values can be found in Appendix~\ref{app}). The stacking of the Fe overlayers starts with layer '1' at the interface with the Cu substrate and continues towards vacuum.}
\begin{ruledtabular}
\begin{tabular}{lcccccc}
System & \multicolumn{4}{c}{exchange parameters (mRy)} & $m_i$ ($\mu_B$/atom) \\
layers (N) & $layer~ indices$ & \multicolumn{1}{c}{inter-layer} & $layer~ index$ & intra-layer &  \\
\hline
$Fe_N/Cu(001)$ & & & & \\
1 &-- &-- & 1&  2.893 & 2.679 & \\
\hline
2 & (1-2) & 0.974 & 1&  2.029 & 2.34 & \\
 & -- & --                   & 2 &  1.789 & 2.91 & \\
\hline
3 & (1-2)& 0.309 & 1&  1.789 & 2.507  \\
   & (2-3)& 0.868 & 2 &  1.760 & 2.354 \\
  & --& -- & 3&   2.039 & 2.962  \\
\hline
$Fe_N/W(110)$ & & & & \\
2 & (1-2) &1.074 & 1&  1.032 & 2.421 \\
&-- & -- & 2& 0.956 & 3.031 \\
\end{tabular}
\end{ruledtabular}
\label{tableFe}
\end{table}
The Fe on Cu(001) overlayers have an easy magnetization axis parallel to the surface normal, with an anisotropy constant of 13.2 $\mu$Ry/layer, while the Fe on W(110) system has an easy magnetization axis along the (1$\overline1$0) direction and an anisotropy constant of 0.3 mRy/layer. 
\begin{table}[hb]
\caption{Exchange parameters and magnetic moments as calculated from first principles, for the Co multilayers systems. All the exchange interactions up to a distance of three lattice parameters have been calculated and included in simulations. For clarity, we list only the dominant nearest neighbour interactions in the table (all the other exchange paramters values can be found in Appendix~\ref{app}). The stacking of the Co overlayers starts with layer '1' at the interface with the Cu substrates (layer '8' being the Co layer at the interface with the vacuum).}
\begin{ruledtabular}
\begin{tabular}{lccccc}
System & \multicolumn{4}{c}{exchange parameters (mRy)} & $m_i$ ($\mu_B$/atom) \\
layers (N) & $layer~indices$ & \multicolumn{1}{c}{inter-layer} &$layer~index$ & intra-layer & \\
\hline
$Co_N/Cu(111)$ &  & & & \\
1 &--& -- & 1& 1.0602 & 1.658 \\
\hline
$Co_N/Cu(001)$ & & & & \\
1&-- & -- & 1 &1.892 & 1.768  \\
\hline
2 & (1-2) & 2.284 & 1& 0.626 & 1.7164  \\
& --& -- & 2& 0.584 & 1.982 \\
\hline
3 & (1-2)& 0.998 & 1& 1.225 & 1.775  \\
& (2-3)& 1.752 & 2&0.337 & 1.772   \\
& --& -- & 3& 1.260 & 1.960  \\
\hline
8 & (1-2) & 1.170 & 1 &1.046 & 1.6122 \\
& (2-3) &1.096 & 2& 0.571 &1.7401  \\
& (3-4) &1.106 & 3& 0.887 &1.7761 \\
& (4-5) &1.133 & 4& 0.79 &1.7689  \\
& (5-6) &1.113 & 5& 0.746 &1.7614  \\
& (6-7) &1.089 & 6& 0.921 &1.7780 \\
& (7-8) &1.532 & 7& 0.486 &1.7231  \\
&--& -- & 8 &1.236 &1.8196 \\
\end{tabular}
\end{ruledtabular}
\label{tableCo}
\end{table}
All the Co systems have an easy magnetization plane which coincides with the plane of the magnetic overlayer. 
The Co/Cu(001) systems have an easy plane along (001) with an anisotropy value of 29.4 $\mu$Ry/layer, while the Co/Cu(111) systems have the easy plane along (111) with an anisotropy constant of 40.7 $\mu$Ry/layer.

For the ferromagnetic systems investigated in this work, the calculated Dzyaloshiskii-Moriya interactions are small, except for the Fe/W(110) system where the DM vectors are significantly large. This is the reason we included these values in Table~\ref{DMFe} in a following section where we discuss in detail the magnon dispersion curve for this system (Sect.~\ref{FeW}).

\subsection{Critical temperatures}
Within the Heisenberg model and in the case where the anisotropy and finite size effects are absent, a two dimensional magnetic film cannot sustain long range magnetic order at finite temperatures (the celebrated Mermin -Wagner theorem). However, the presence of symmetry breaking perturbations, such as small anisotropies, will overrule Mermin-Wagner theorem and finite temperature magnetism is possible, even in the case of a single magnetic layer. The critical temperatures are often reduced in thin films compared to their bulk counterpart. Using our \textit{ab initio} calculated exchange interaction parameters, anisotropy and magnetic moments, we estimate the critical temperature, $T_c$, in several thin magnetic films by means of Monte Carlo simulations and using the cumulant crossing method~\cite{LandauMC}. The systems we investigated are Co$_N$/Cu(111), Co$_N$/Cu(001), Fe$_N$/Cu(001) and Fe$_N$/W(110), where $N$ indicates the number of magnetic atomic layers present. All these systems have ferromagnetic configuration as the magnetic ground state. The results are summarized in Table~\ref{tab:cocu_tc}. It is difficult to directly compare the calculated $T_c$ values with experimental results\cite{Vaz2008,Taroni2008}, due to the lack of information about interface quality, inter-diffusion between the substrate and the magnetic film and the surface roughness of the samples. All these imperfections affect the value of the critical temperature.~\cite{Holmstrom:2004fk} The calculations assume the idealized case with perfect interfaces. For the considered systems, with non-magnetic Cu and W substrates, the presence of perfect interfaces is likely to give the highest possible value of $T_c$ that the system can reach. Our data for 3 ML Co/Cu(111) are in rather good agreement with the measured values for 2.5 ML Co/Cu(111), although a comparison should be done with care, since the experiment and theory have slightly different Co thicknesses. For the 1ML Co/Cu(111) system as well as for the bulk Co, the theoretical values seem to be in agreement with measurements, whereas for the 2ML Co/Cu(001) system the calculated value of $T_c$ is two times larger than the measured value. For 2ML Fe/W(110) as well as for bulk Fe, theory and experiments are in good agreement, whereas for the Fe/Cu(001) system, theory consistently overestimates the measured values. 
Our calculations do however agree reasonably well with the results obtained by Pajda {\it et. al}~\cite{Pajda2000} using the random phase approximation (RPA).
%
\begin{table}[h]
\caption{Calculated values of the Curie temperatures $T_C$ for Co and Fe overlayers on W and Cu substrates with different orientations and compared with published values determined theoretically and experimentally.}
\begin{ruledtabular}
\begin{tabular}{lrcl}
\multicolumn{1}{c}{} &
\multicolumn{2}{c}{T$_C$ (K)} \\
\multicolumn{1}{c}{\bf System} &
\textrm{{\bf Theory:} this work} &
\textrm{previous studies} &
\textrm{\bf Experiment} \\
\colrule 
bulk fcc Co & 1120 & 1080~\cite{Rosengaard1997},1250~\cite{Halilov1998},1311 $\pm$ 4~\cite{Pajda2001} & 1388-1398 \\
{\it Co$_N$/Cu111} & & & \\
1    & 255 & $-$ & 207-434~\cite{Vaz2008} \\
2    & 660 & $-$ & 500 [1.7 ML]~\cite{Vaz2008} \\
{\it Co$_N$/Cu001} & & & \\  
1    & 370 & 426~\cite{Pajda2000} & $-$ \\
2    & 680 & $-$ & 325~\cite{Vaz2008} \\
3    & 705 & $-$ & 650 [2.5 ML]~\cite{Huang:1994fk} \\
8    & 1070 & $-$ & $-$ \\
{\it Fe$_N$/Cu001 } & & \\
1    & 400 & 515~\cite{Pajda2000} & $\approx$100~\cite{Vaz2008} \\ 
2    & 380 & $-$  & $\approx$270~\cite{Vaz2008} \\
3    & 560 & $-$ & $\approx$290~\cite{Vaz2008} \\
{\it Fe$_N$/W110} & & &\\
2    & 520 & $-$ & 450~\cite{Elmers1995} \\
\end{tabular}
\end{ruledtabular}
\label{tab:cocu_tc}
\end{table}
\vfill
\subsection{Spin wave spectra}
\subsubsection{ {\bf Co on Cu(001)}}
The fcc Co/Cu(001) system represents a model system for the study of magnetic phenomena in thin films, since it does not exhibit strong structural, chemical or magnetic instabilities.~\cite{Gradmann1993,Vaz2008} It was therefore a natural candidate for investigation with SPEELS, and became the first ultrathin system in which its magnon spectrum was measured using this method.~\cite{Vollmer2003} The system could be described to a good level of accuracy within the context of the nearest-neighbour Heisenberg model on a semi-infinite substrate. In this case a surface mode exists with a dispersion curve $\hbar\omega^{\mathrm{(surf)}} = 8JS(1-\cos(qa_0))$ along the $\langle 110 \rangle$ direction, where $J$ is the exchange coupling, $S$ is the magnitude of the spin per primitive unit cell, $q$ is the length of the magnon wave vector and $a_0 = 2.55$ \AA{}. The fit of this curve to the measured SPEELS data gave $JS=15.0 \pm 0.1$ meV, which compares well with the value of $JS=14.7 \pm 1.5$ meV obtained from neutron data of bulk fcc Co at long wave lengths~\cite{Sinclair1960} (\emph{i.e.} in the regime $q<0.3$ \AA{}$^{-1}$). 

The Halle group followed up their initial report for the 8 ML film~\cite{Vollmer2003} with results relevant to Co/Cu(001) systems with decreasing thickness down to 2.5 ML.~\cite{Etzkorn2004,Vollmer2004b} These experiments demonstrated a very weak reduction in the energies required to excite the spin waves, relative to the bulk.~\cite{Sinclair1960} However, for all the thicknesses reported, the surface mode at the surface Brillouin zone boundary (the $\mathrm{\overline{X}}$ point) is well below the bulk band edge, at around 240 meV.  This difference is for the most part caused by the reduced number of nearest neighbours (NN) at the surface (only 8 atoms) with respect to the bulk case, where there are 12 NN present.~\cite{Vollmer2003} For energy ranges where Stoner excitations are important, no direct comparison between calculated and experimental data for bulk Co is possible at the moment.

Theoretical investigations of Co/Cu(001) thin films have been carried out using both the adiabatic approximation~\cite{Udvardi2003,Pajda2000} and the random phase approximation (RPA) to a description of the spin response of the itinerant electron system.~\cite{Costa2004a,Costa2004b} The primary difference between the two approaches concerns the treatment of the particle hole excitations known as Stoner excitations,~\cite{Mohn2003} which are neglected in the adiabatic case. Broadly speaking, these are not relevant at low energies, where the density of magnon states dominates. However, at higher energies or large wave-vectors their role becomes significant, and one would therefore expect theories based on the adiabatic approach to break down at the Brillouin zone boundary. In addition, Stoner excitations are more pronounced for low dimensional magnets compared to bulk magnets.

\begin{figure}[htp]
     \centering
    \begin{tabular}{cc}
   \hspace{-1cm}
     \subfigure[]{
          \label{fig1a}
          \includegraphics[scale=0.3]{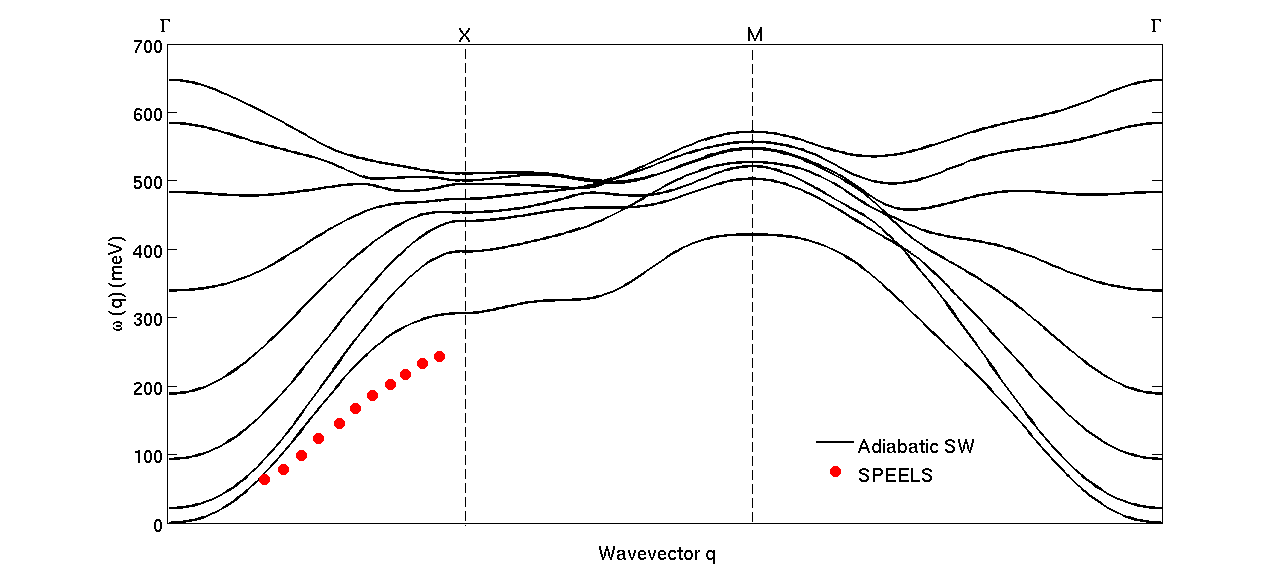}}
     \hspace{-2cm}
     \subfigure[]{
          \label{fig1b}
          \includegraphics[scale=0.3]{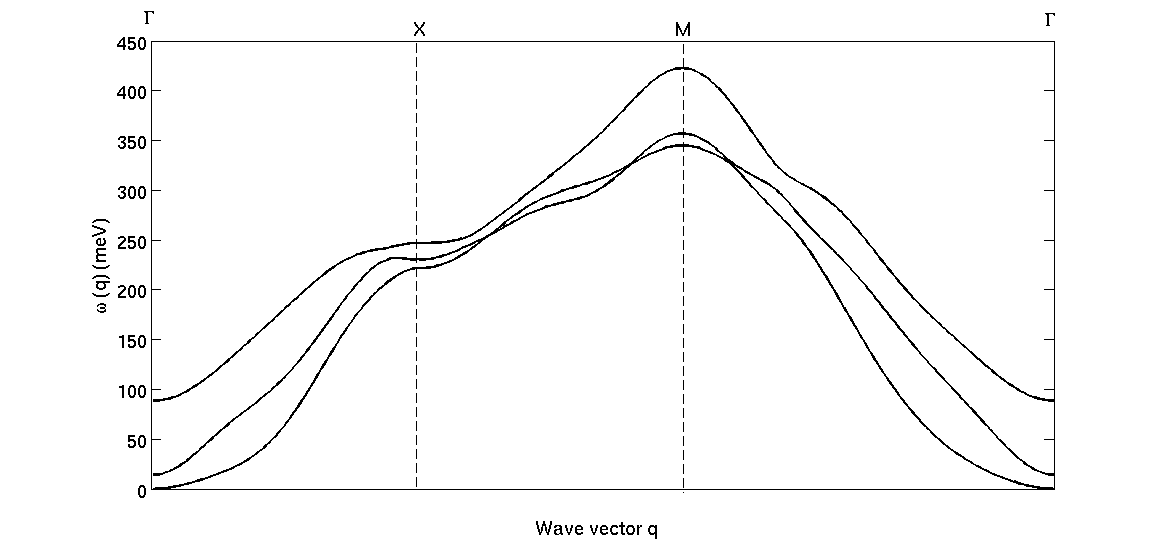}}
  \end{tabular}
     \caption{\label{fig1}: (Color online) Adiabatic magnon spectra (full lines) obtained from SPR-KKR calculations for (a) 8 ML Co/Cu(001) together with experimental SPEELS data~\cite{Vollmer2003} (circles) and (b) 3 ML Fe/Cu(001).  }
\end{figure}

Fig.~\ref{fig1a} displays the calculated adiabatic magnon spectrum obtained for 8 ML Co/Cu(001), using Eq.~(\ref{eqn:Jijmatrix}) with $J_{ij}$ values obtained using the LKG method.~\cite{Liechtenstein1984,Liechtenstein1986} The most notable feature is the presence of several branches, one for each Co layer present. This is in contrast to experimental observations, where only the lowest ("acoustic'') branch~\cite{Vollmer2003} (filled circles in Fig~\ref{fig1a}) as well as the second lowest branch ~\cite{Costa2006} are observed. As Vollmer \emph{et al.} point out,~\cite{Vollmer2003} this indicates the shortcomings of a direct interpretation of their data in terms of a na\"{i}ve Heisenberg model. 

Fig.~\ref{fig1b} displays the adiabatic magnon spectrum obtained for 3 ML Fe/Cu(001). Unfortunately, there are no experimental data to compare with. The values for the spin wave stiffness, $A$, can be estimated from the calculated adiabatic magnon spectra by measuring the curvature of the dispersions as $\mathbf{q}\rightarrow 0$. For this system we find $A$ being approximately 210 meV \AA$^2$. For the 1 ML Co on Cu(001) case we obtain a value of the order of 420 meV \AA{}$^2$ - a 15 \% overestimate of the experimentally determined value of 360 meV \AA{}$^2$, but considerably softer than the theoretical value determined by Pajda \emph{et al.} of 532 $\pm$ 9~meV \AA{}$^2$,~\cite{Pajda2000} obtained by a real-space adiabatic approach. For 1 ML Fe on Cu(001) we obtain a lower value, of 260 meV \AA{}$^2$ compared to the value of 331 meV \AA{}$^2$ by Pajda \emph{et al.}~\cite{Pajda2000}.
           
As explained above, the adiabatic approximation becomes questionable at higher energies and/or large wave-vecors, since Stoner excitations become relevant. In order to address this issue, Costa \emph{et al.}~\cite{Costa2004a,Costa2004b} have developed a theory that explicitly takes these excitations into account. Their approach successfully describes the SPEELS measurements for the 8 ML Co/Cu(001) system: their calculated magnon spectrum is in agreement with the experiment over all the $\overline{\Gamma}-\overline{\mathrm{X}}$ line and correctly predicts a broadening of the ``acoustic'' spin wave peaks, along with an absence of standing spin waves giving rise to ``optical'' branches. Their work strongly indicates that a process known as Landau damping, through which the spin waves decay into the Stoner continuum, is at play in the Co/Cu(001) system.

At present, the ASD method cannot handle one-particle Stoner excitations directly (these are responsible for the longitudinal fluctuations of the magnetic moments), although recently efforts have been made to include these effects in the LLG-equation, both directly\cite{Dudarev2012} and indirectly\cite{Chimata2012}. However, the ASD method allows to address an alternative source of damping, namely the temperature, and we have performed simulations for different thicknesses of Co overlayers (1, 2, 3 and 8 atomic layers) under two very different conditions (Figs.~\ref{fig2} -- \ref{fig6}). First we performed simulations at low temperature (1~K), where the damping constant $\alpha$ in Eq.~(\ref{eqn:LLG}) is set to 3 $\times$ 10 $^{-4}$, a value that ensures a very weak coupling to the temperature bath. In this case the temperature effects are deliberately kept to a minimum. Secondly, we perform simulations in more realistic conditions, namely at room temperature (T=300~K) and a physically more plausible damping constant $\alpha=0.05$ with the exception of 1 ML on Cu(001) and Cu(111) where temperature was set to 200 K to make sure the simulation conditions were below the Curie temperature. All these results are discussed in detail in the following sections.

\subsubsection*{1 ML}

\begin{figure}[h] 
     \begin{tabular}{cc}
     \hspace{-1cm}
     \subfigure[]{
          \label{fig2a}
          \includegraphics[scale=0.40,angle=270]{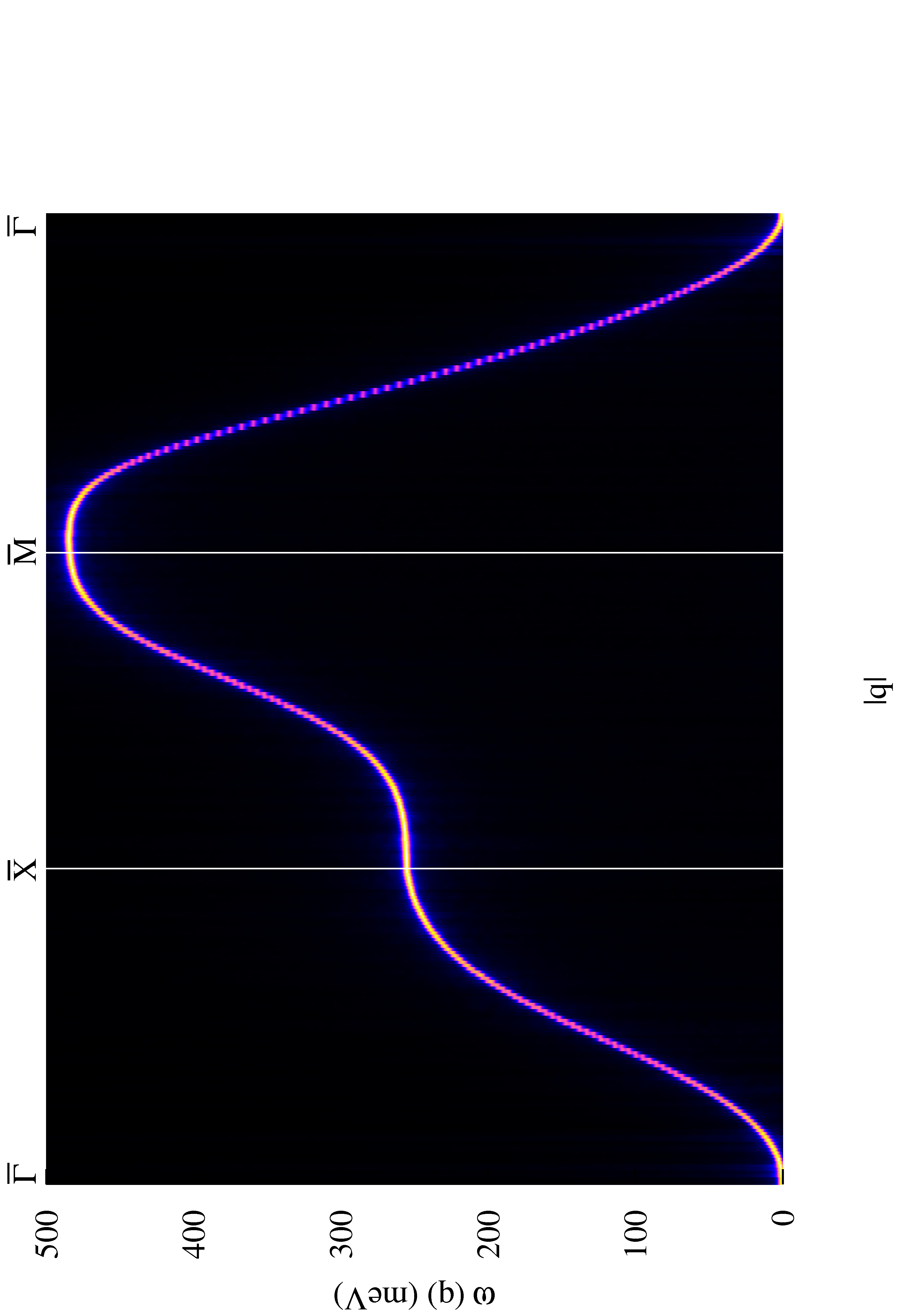}}
     \hspace{-1cm}
     \subfigure[]{
          \label{fig2b}
          \includegraphics[scale=0.40,angle=270]{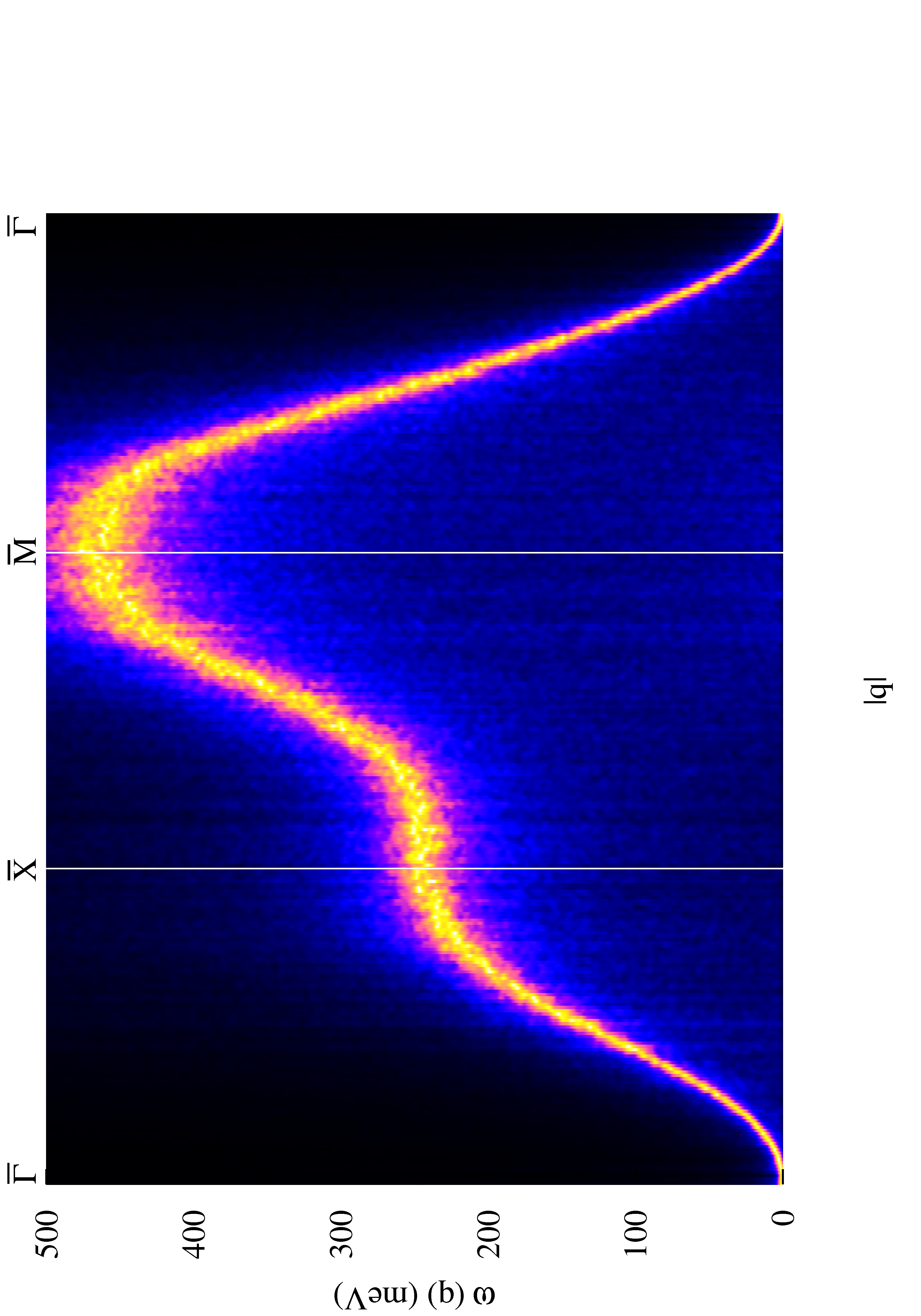}}
      \end{tabular}
     \caption{\label{fig2} (Color online) Spin wave dispersion spectra obtained from ASD simulations of 1 ML Co/Cu(001) at (a) T= 1K and small damping constant $\alpha=3$ $\times$ 10 $^{-4}$ and (b) T = 200 K and realistic damping constant $\alpha=0.05$ }
\end{figure}

Fig.~\ref{fig2} displays the spin wave spectra in a 1 ML thick Co layer on Cu(001). The values of the exchange parameters In the low temperature case (Fig.~\ref{fig2a}), the spectra obtained from the atomistic spin dynamics simulations is almost indistinguishable from the adiabatic spectra. We point out that these results are obtained from a calculation of the dynamical structure factor $S(\mathbf{q},\omega)$. 
In order to facilitate the comparison between the spin wave spectra for a single layer of Co on Cu for different orientations of the substrate, we performed the calculations for both Cu(001) and Cu(111) orientations (discussed in detailed below), at the same value of the temperature, i.e. T= 200~K. 
 Increasing the temperature and damping (Fig.~\ref{fig2b}) causes some broadening of the dynamical structure factor and as a result the magnon energy at the BZ boundary ($\mathrm{\overline X}$ and $\mathrm{\overline M}$ points) decreases slightly. However, overall the effects of temperature and dynamic treatment are not significant in this case. 

\subsubsection*{2 and 3 ML}

Qualitatively, the 2 ML and 3 ML case are not so different so we only display the calculated dynamic structure factor for 2 ML Co in Figure 3.%
\begin{figure}[h]
\begin{tabular}{cc}
\hspace{-1cm}
     \subfigure[]{
          \label{fig3a}
          \includegraphics[scale=0.40,angle=270]{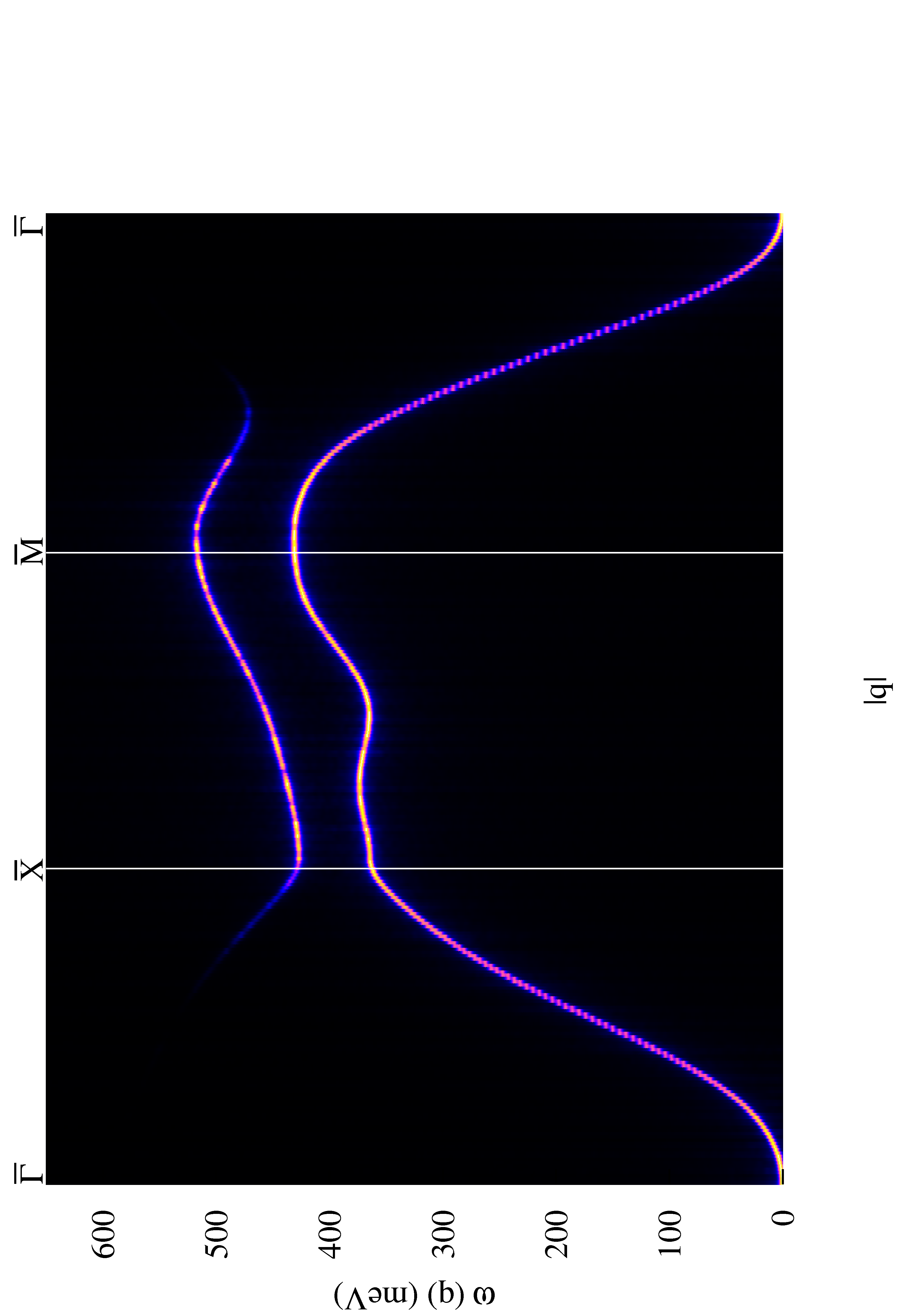}}
     \hspace{-1cm}
     \subfigure[]{
          \label{fig3b}
          \includegraphics[scale=0.40,angle=270]{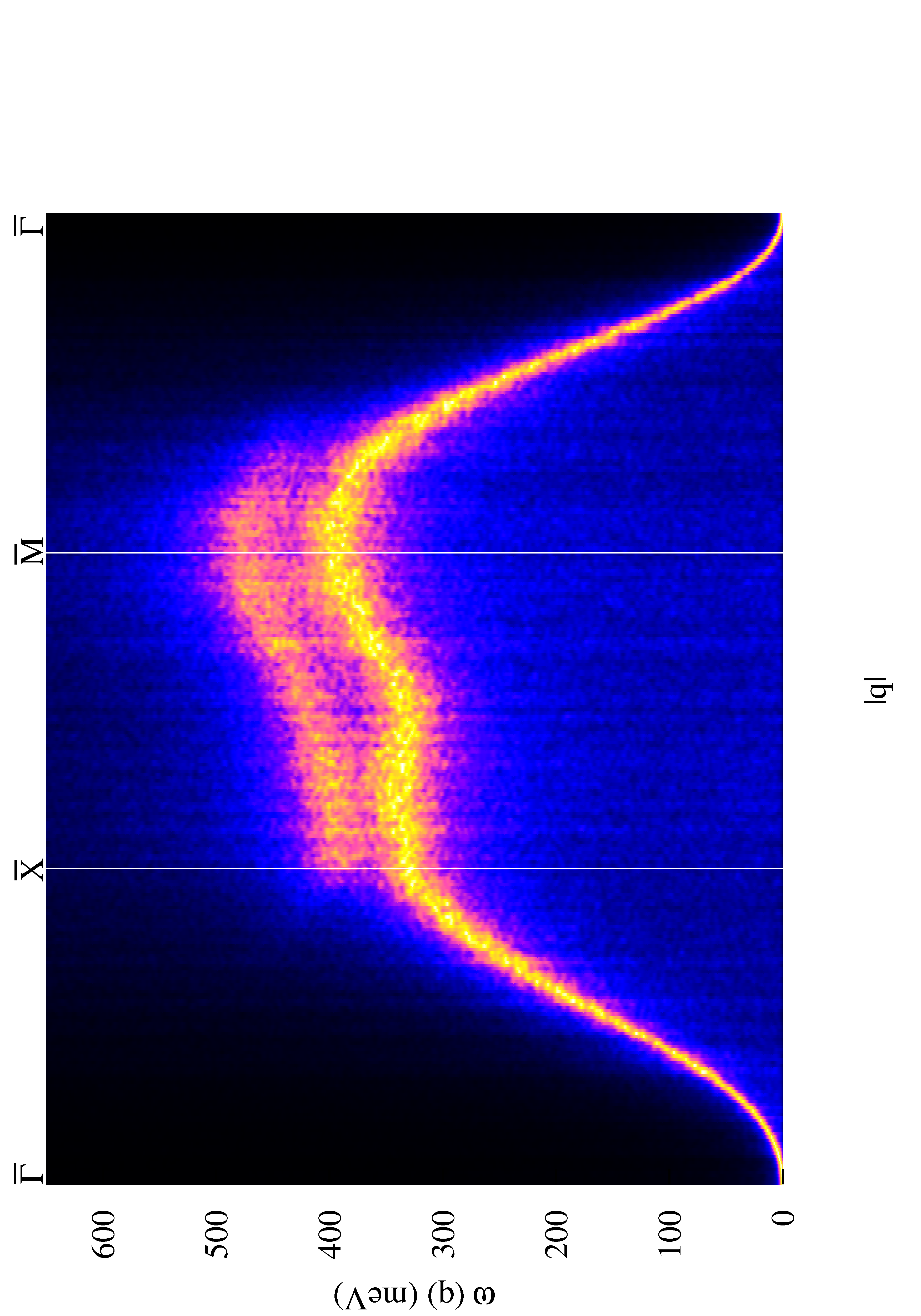}}
\end{tabular}  
     \caption{\label{fig3} (Color online) Spin wave dispersion spectra  obtained from ASD simulations of 2 ML Co/Cu(001) at a) T= 1K and small damping constant $\alpha=3$ $\times$ 10 $^{-4}$ and b) T = 300 K and realistic damping constant $\alpha=0.05$ }
\end{figure}
Having more than one layer on top of Cu(001), we expect, apart from the acoustic branch, a number of optical branches to appear in the magnon spectra, similar to what is found in the adiabatic spectra. In the case of 2 ML we expect one branch of each kind, but as noticed in Fig.~\ref{fig3a}, even at very low temperature (T= 1~K) and extremely small damping, the optical branch is very weak, especially at small wave vectors, close to the $\mathrm{\overline \Gamma}$-point. In Section~\ref{sec:SQanalysis} below, we make a deeper analysis of the dynamical structure factor to better understand the suppression of the optical branches. This type of analysis has been recently published by Taroni et al..~\cite{Taroni2011} Using a more reasonable temperature and damping (Fig.~\ref{fig3b}) erases most traces of the optical branch which is smeared out and remains slightly visible only in the $\mathrm{\overline X - \overline M}$ region. The magnon energy at the BZ boundaries is reduced in this case by roughly 25 meV which can be attributed to temperature effects.  

\subsubsection*{8 ML}

\begin{figure}[h]
\begin{tabular}{cc}
\hspace{-1cm}
     \subfigure[]{
          \label{fig4a}
          \includegraphics[scale=0.40,angle=270]{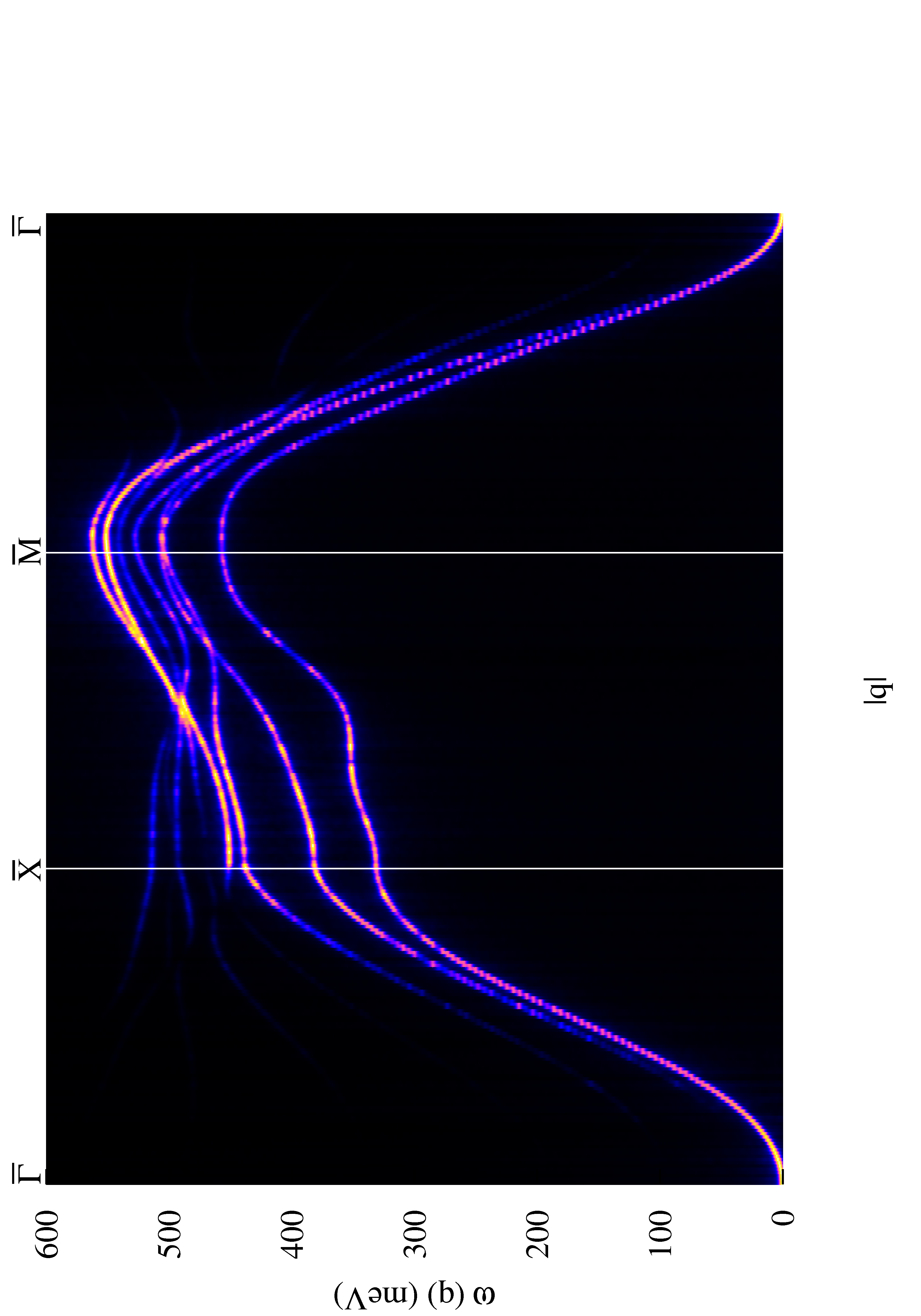}}
     \hspace{-0.8cm}
     \subfigure[]{
          \label{fig4b}
          \includegraphics[scale=0.40,angle=270]{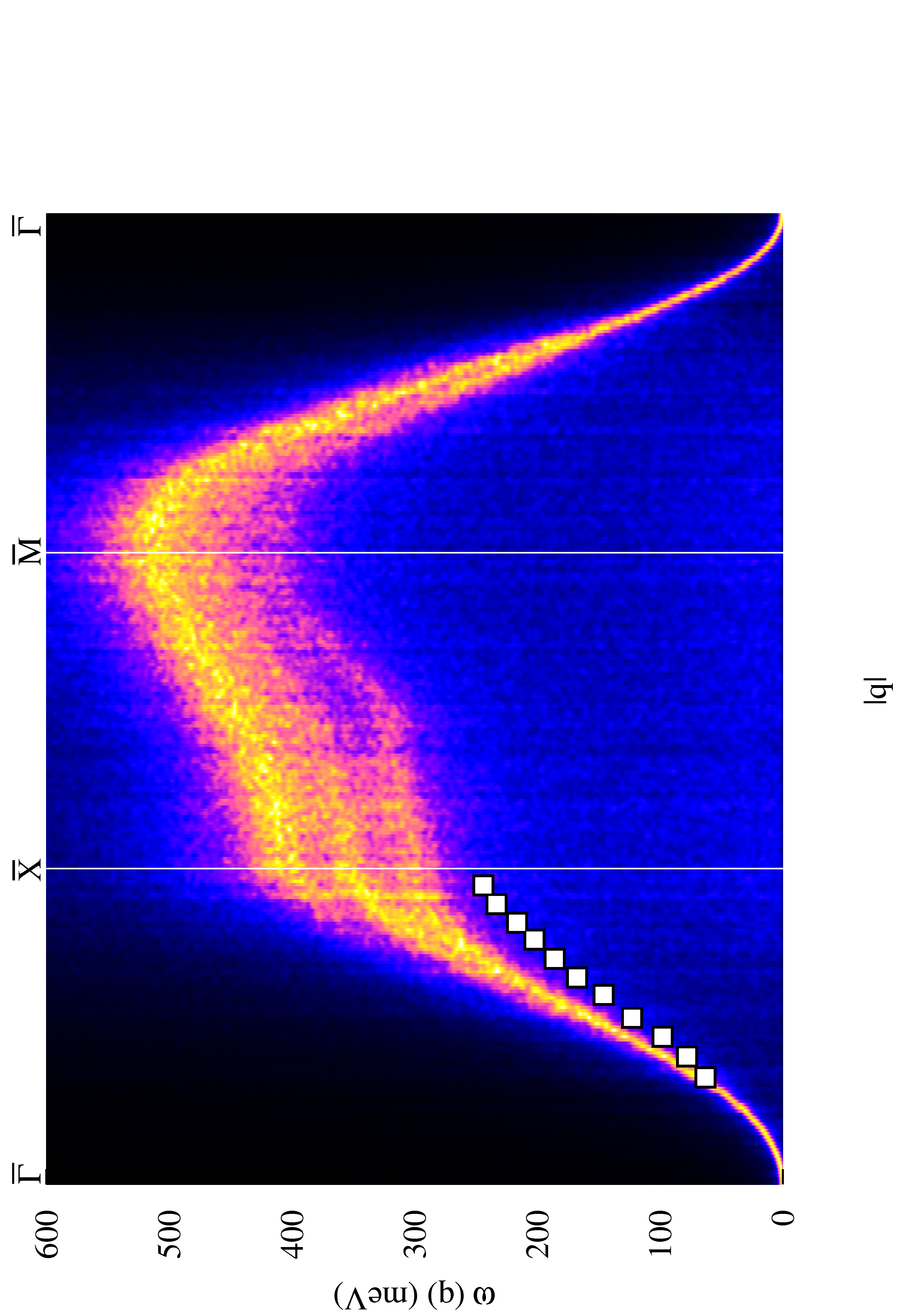}}
\end{tabular}
     \caption{\label{fig4} (Color online) Spin wave dispersion spectra obtained from ASD simulations of 8 ML Co/Cu(001) at a) T= 1K and small damping constant $\alpha=3$ $\times$ 10 $^{-4}$ and b) T = 300 K and realistic damping constant $\alpha=0.05$. The experimental values  are marked by white squares.~\cite{Vollmer2003}}
\end{figure}

Fig.~\ref{fig4} displays the spin wave spectra on the full 8 ML Co stack on Cu(001). It is immediately noticed that the contrast between the calculated spectrum  based on the dynamical structure factor (Fig.~\ref{fig4a}) with the spectrum directly obtained from first-principles (Fig.~\ref{fig1a}) is strong for values of $q$ close to the $\bar \Gamma$-point. In Fig.~\ref{fig4}, there is little trace of the ``optical'' branches even in the case of low temperature and damping. We also note that at 300 K, the magnon energies at the zone boundaries are reduced by roughly 25 meV, similar to the 2 ML case. The suppression of optical branches arising from a dynamical treatment has also been pointed out by Costa \emph{et al.}~\cite{Costa2004a,Costa2004b}. We also note that the agreement between the theoretical acoustic branch that we calculated and experiment is rather good.

\subsubsection{{\bf Co on Cu(111)}}

\begin{figure}[htb]
     \centering
          \label{fig:cocu111}
          \includegraphics[scale=0.40,angle=270]{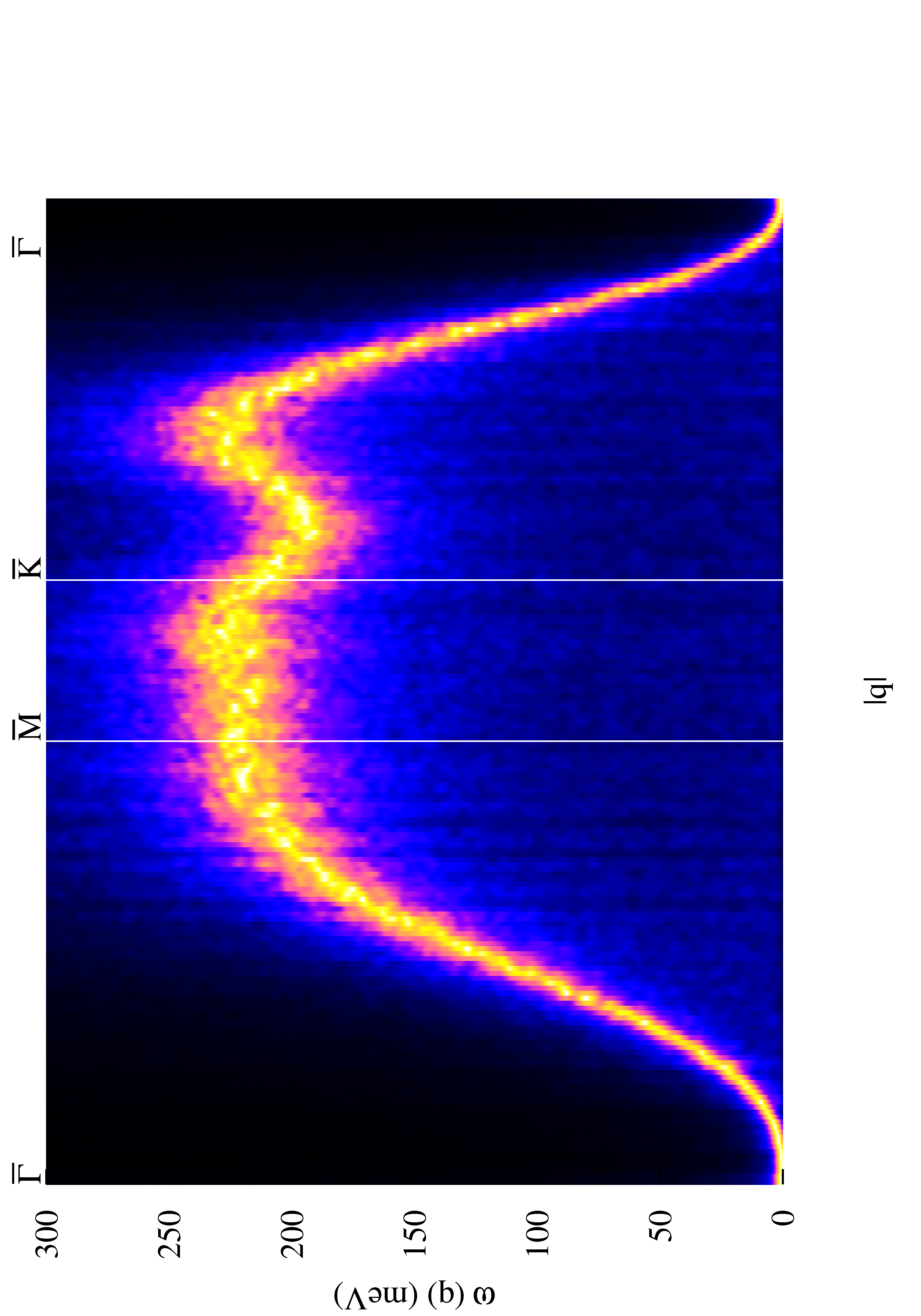}
     \caption{\label{fig5} (Color online) Spin wave dispersion spectra obtained from ASD simulations of 1 ML Co/Cu(111) at T = 200 K and realistic damping constant $\alpha=0.05$ }
\end{figure}

For completeness, we also perfomed simulations on a single layer of Co on the Cu(111) surface. Due to the fact that the critical temperature of the single Co layer on Cu(111) is lower than the room temperature (i.e. T$_c$=255~K), for this system we performed the high-temperature calculations at 200~K. In Fig.~\ref{fig5}, the spin wave spectra obtained at T=200K is displayed. The calculated spin wave spectra is noted to be softer than that of the 001-surface, which is consistent with the lower Curie temperatures obtained for this system. The calculated spin wave stiffness has a value of around 220 meV\AA{}$^2$, almost half of the value compare to what was found for the 001-surface.

\subsubsection{{\bf Fe on Cu(001)}}

\begin{figure}[h]
\begin{tabular}{cc}
\hspace{-1cm}
     \subfigure[]{
          \label{fig6a}
          \includegraphics[scale=0.40,angle=270]{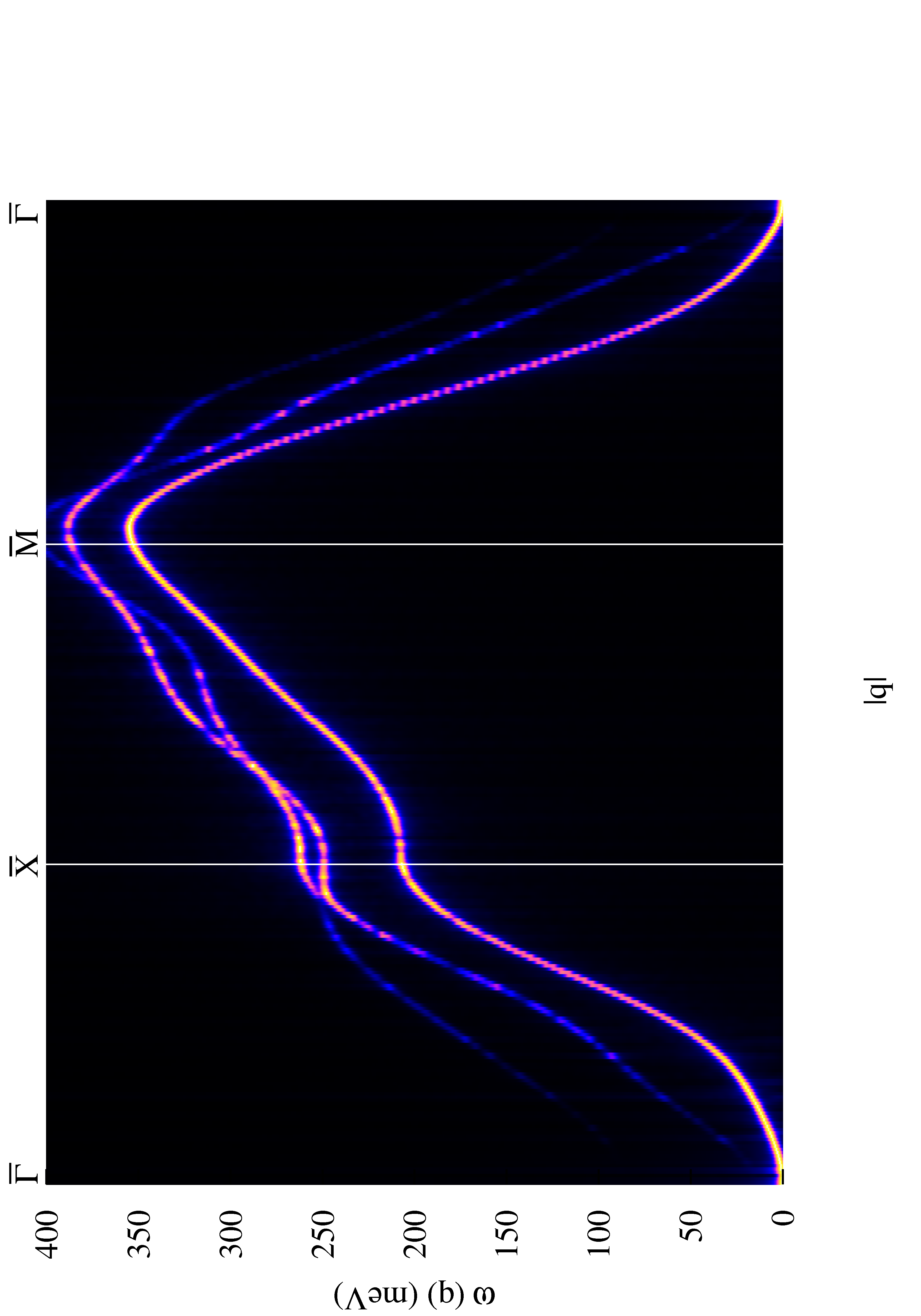}}
     \hspace{-1cm}
     \subfigure[]{
          \label{fig6b}
          \includegraphics[scale=0.40,angle=270]{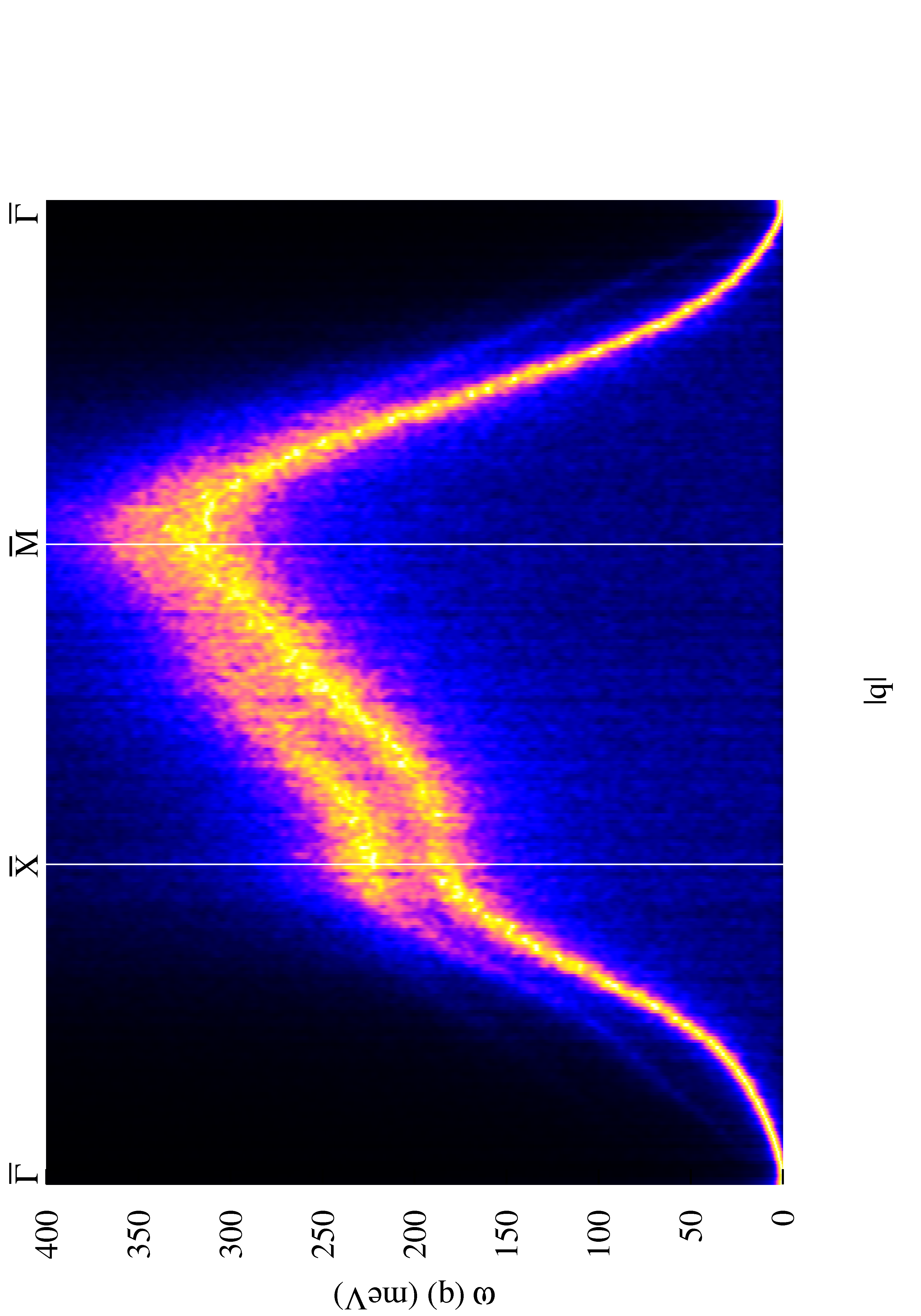}}
\end{tabular}
     \caption{\label{fig6} (Color online) Spin wave dispersion spectra obtained from ASD simulations of 3 ML Fe/Cu(001) at a) T= 1K and small damping constant $\alpha=3$ $\times$ 10 $^{-4}$ and b) T = 300 K and realistic damping constant $\alpha=0.05$ }
\end{figure}

The Fe/Cu(001) system is more complex than Co/Cu(001) for a number of reasons. Since Fe doesn't naturally exist in the fcc phase it is difficult to grow thicker layers of Fe with good quality. It is also well known, from theory, that bulk fcc Fe exhibits a complicated magnetic phase diagram with many magnetic configurations with similar energies. The same  thing is true for thick Fe layers on Cu. However, for thin Fe layers (1 to 3 ML), it is generally accepted that Fe adopts a ferromagnetic configuration. In the case of thicker layers there are several proposed magnetic configurations, for instance Sandratskii\cite{SandratskiiFeCu} claimed that the magnetic structure takes the form $\downarrow \uparrow \uparrow$ for the three upper layers. We performed Monte Carlo simulations using exchange parameters starting from either ferromagnetic or the proposed magnetic structure of Sandratskii as reference state for the 3 ML case. Regardless of the starting configuration, we always obtained the ferromagnetic configuration as the ground state magnetic structure also for 3 ML Fe on Cu001. However, the spin waves (Fig.~\ref{fig6}) are soft which is reflected in the calculated adiabatic spin wave stiffness constant $A$, which is considerably lower than what we obtain for 1 ML. We would like to point out that in general ASD gives a more realistic description of the magnon spectra than the adiabatic approximation (Fig.~\ref{fig1b}).

In Fig.~\ref{fig6} we show the results of spin wave spectra only for the 3 ML case. The spectra is rather different from Co/Cu(001) and the most striking difference is the overall softness of the magnons. However, we should keep in mind that the critical temperature for the 3 ML Fe/Cu(001) is much lower than for the analogous Co system.
The optical branch for the Fe system is more pronounced at low temperatures but similar to the Co case it is suppressed close to the $\bar \Gamma$-point for reciprocal vectors inside the Brillouin zone. For realistic conditions, i.e. at room temperature and for larger damping, the intensity of the optical branches is very weak and smeared traces of those branches remains visible mostly in the $\mathrm{\overline X - \overline M}$ region. 

\subsubsection{{\bf Fe on W(110)}}
\label{FeW}

\begin{figure}[h]
\begin{tabular}{cc}
\hspace{-1cm}
     \subfigure[]{
          \label{fig7a}
          \includegraphics[scale=0.40,angle=270]{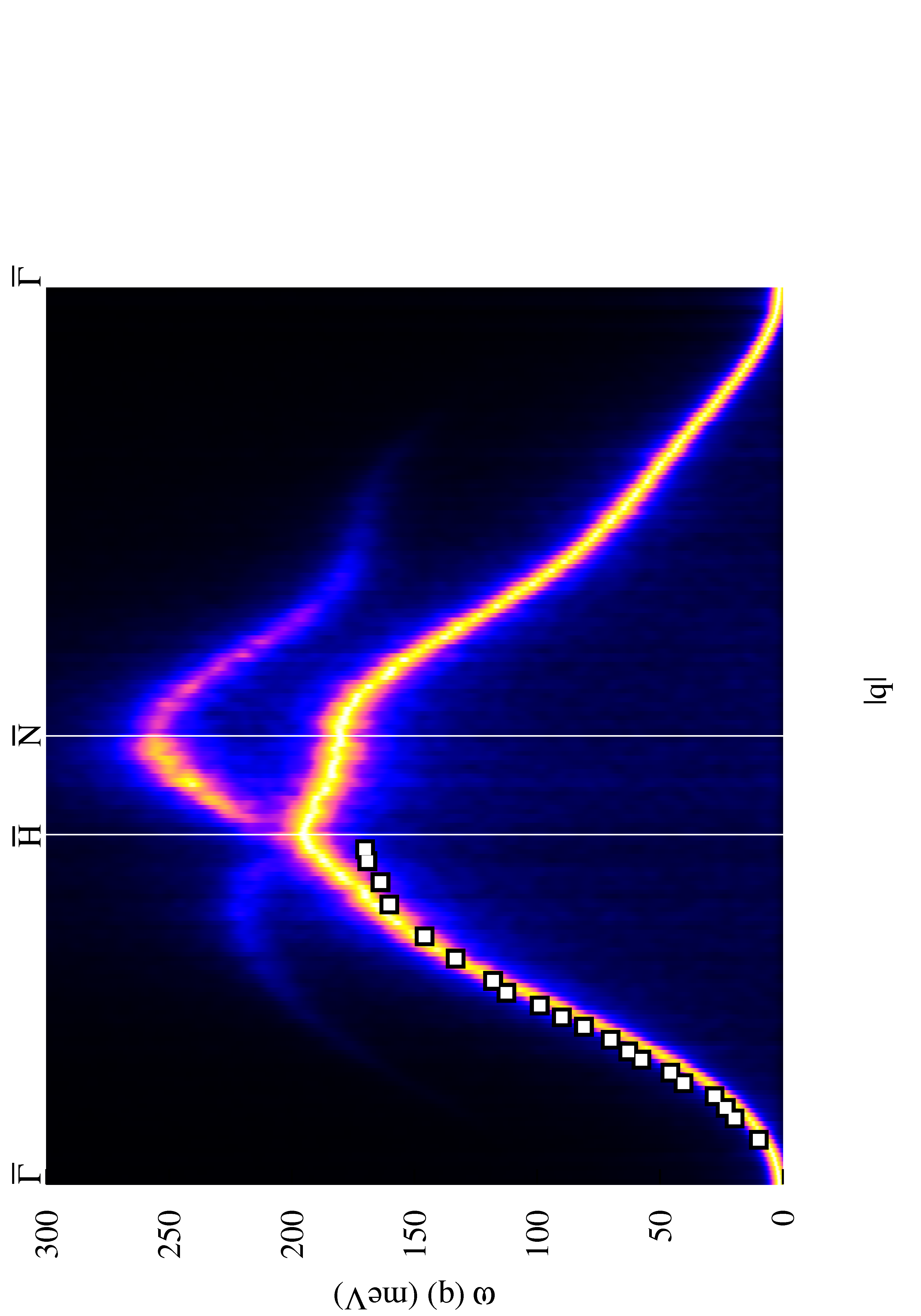}}
     \hspace{-1cm}
     \subfigure[]{
          \label{fig7b}
          \includegraphics[scale=0.40,angle=270]{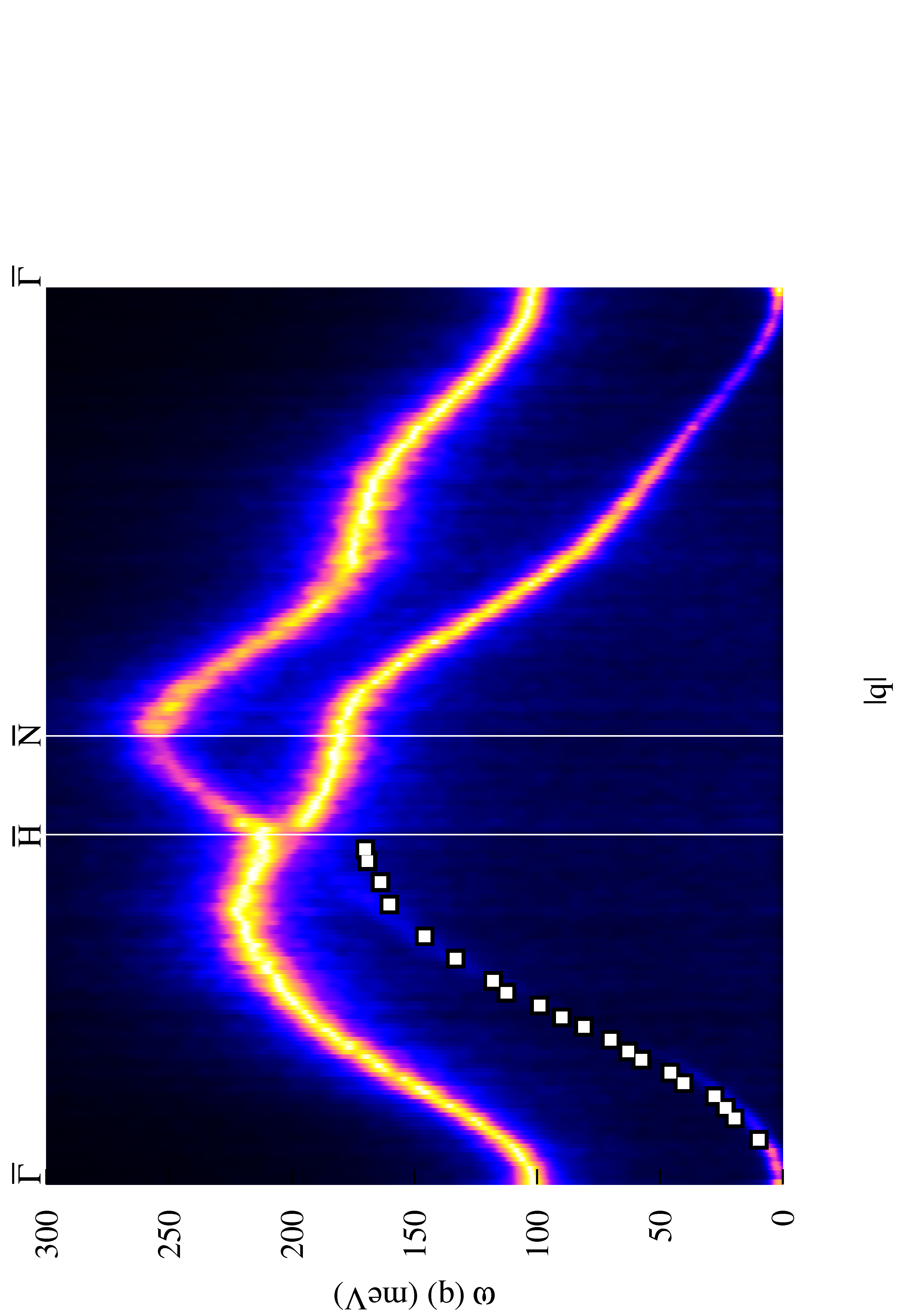}}
\end{tabular}
     \caption{\label{fig7} (Color online) Spin wave dispersion spectra obtained from ASD simulations of 2 ML Fe/W(110) at T=300 K and $\alpha=0.01$. a) Sampling inside the first Brilloiun zone ($\boldsymbol{\tau}=[0 0]$) b) Sampling shifted by vector $\boldsymbol{\tau}=[1 0]$. Experimental values are marked by white squares.~\cite{Tang2007}. See the text for details.}
\end{figure}

In a previous study~\cite{Bergman2010}, we performed calculations of the spin wave spectra of a single Fe layer on top of W(110) and compared it with the experimentally determined spectra~\cite{Prokop2009} and found a good agreement. In particular, we observed a softening of the magnons with respect to the bulk spin wave dispersion spectra. This magnon softening in thin layers is expected even when considering a simple NN Heisenberg model for the description of the dispersion curves in ferromagnets, where the exchange parameters are weaker due to a lower coordination at the surface. Here we focus on the magnons of a Fe bilayer on W(110). Experimentally, the magnon dispersion has been measured by SPEELS~\cite{Tang2007} where in addition the magnon lifetimes were determined~\cite{Zakeri2012}. In Fig.~\ref{fig7} we show our calculated magnon dispersion along symmetry lines in the first Brilloiun zone together with experimental data. Along the line $\bar \Gamma - \bar H$, we find a very good agreement between experiment and theory (Fig.~\ref{fig7a}). If the momentum transfer is changed, so that one goes outside the first Brillouin zone, the different spin wave branches can be determined. This is analyzed in detail in Section~\ref{sec:SQanalysis}. In Fig.~\ref{fig7b} we show the so obtained magnon dispersion, where the optical modes can be clearly observed.

\begin{figure}
  \includegraphics[width=9.5cm,angle=270]{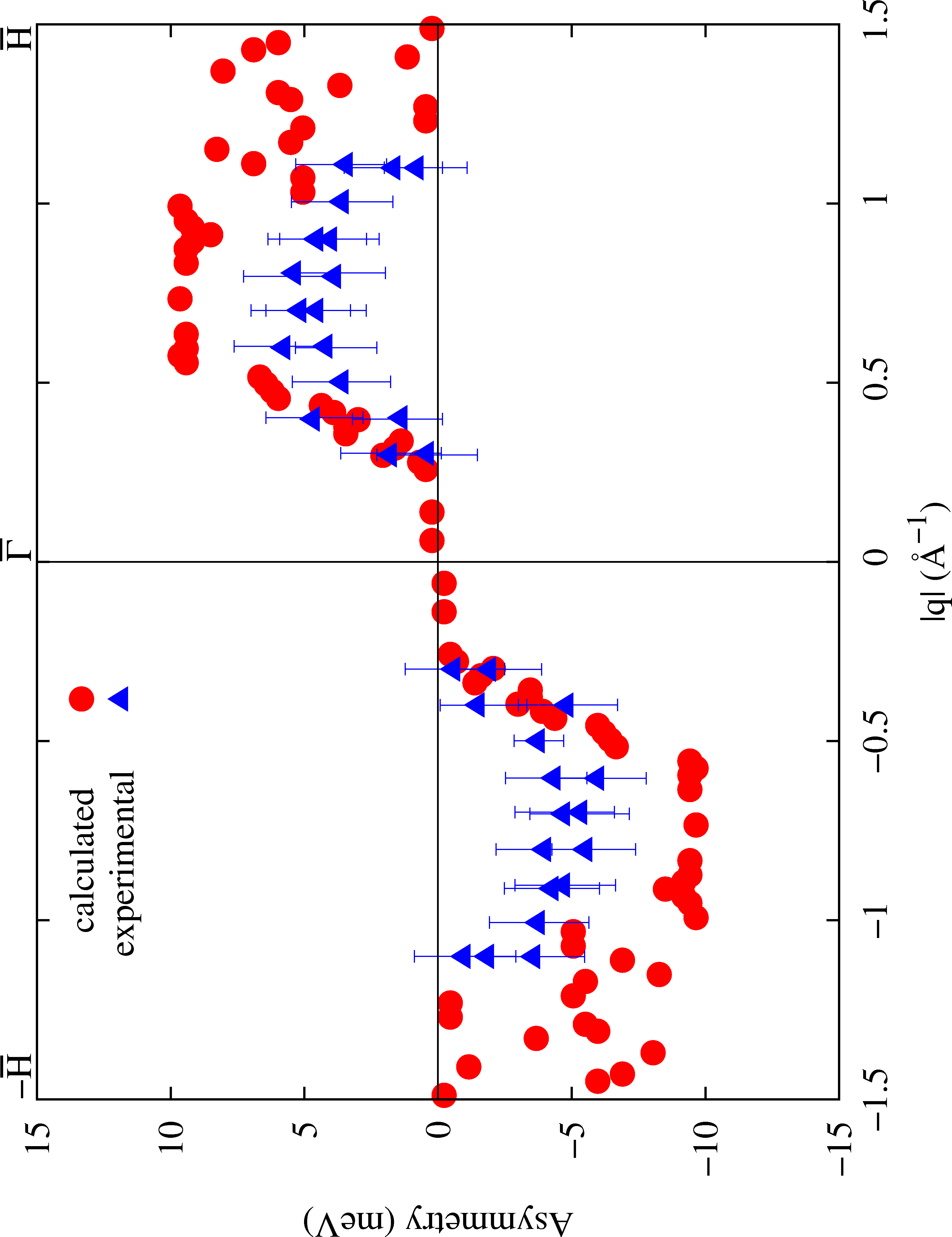}
  \caption{\label{fig8} (Color online) Calculated spin wave asymmetry for the magnon spectrum of 2 ML Fe/W(110), using theoretically determined Dzyaloshinskii-Moriya interactions. The experimental values have been obtained by Zakeri et al.~\cite{Zakeri2010} for M $\parallel [\bar 110]$.}
\end{figure}

 Udvardi and Szunyogh~\cite{Udvardi2009} predicted the asymmetry in the magnon spectrum arising from the Dzyaloshinskii-Moriya interaction. Later on it was experimentally detected by Zakeri \emph{et al.}~\cite{Zakeri2010} in a Fe bilayer on W(110). The spin wave asymmetry $\Delta E$ is defined as the difference in the spin wave energy $\omega(\mathbf{q})$ between $\mathbf{q}$ and $-\mathbf{q}$, i.e $\Delta E= \omega(\mathbf{q})-\omega(-\mathbf{q})$. If Dzyaloshinskii-Moriya interactions are absent, i.e. very weak effect from spin-orbit interaction, the asymmetry is zero for every wave vector $\mathbf{q}$. In Fig.~\ref{fig8}, we show our calculated spin wave asymmetry for the Fe bilayer on W(110) for wave vectors ranging from $-\bar{H}$ to $\bar{H}$ in the two dimensional Brillouin zone, using theoretically determined Dzyaloshinskii-Moriya interaction parameters, see Table IV. 

\begin{table}
\caption{Dzyaloshinskii-Moriya interactions as calculated from first principles, for the Fe bilayer on W(110). All the Dzyaloshinskii-Moriya vectors up to a distance of three lattice parameters have been calculated and included in simulations, but for clarity we list only the $x$ and $y$ component of the first two shells in the table (the $z$ component is zero). The interface Fe layer has label '1' and surface layer '2'. }
\begin{ruledtabular}
\begin{tabular}{lcc}
Type of interaction & \multicolumn{2}{c}{Dzyaloshinskii-Moriya interactions (mRy)}  \\
 & \ $| D_x | $  & $ | D_y | $  \\
\hline
intra-layer & & \\
(1-1) & 0.038 & 0.069  \\
      & 0.000 & 0.085  \\
(2-2) & 0.012 & 0.021  \\
      & 0.000 & 0.027  \\
inter-layer  & & \\
(1-2) & 0.000 & 0.051 \\
      & 0.026 & 0.000 \\
\end{tabular}
\end{ruledtabular}
\label{DMFe}
\end{table}

 The simulations were performed at room temperature, as in experiment, with realistic damping. We obtain a qualitatively good agreement with experiment but the amplitude of the calculated spin wave asymmetry is slightly overestimated ( $\approx 12 meV$ compared to $\approx 8 meV$ in experiment). There are several explanations to this discrepency, primarily the asymmetry is sensitive to the value of the Dzyaloshinskii-Moriya interaction which is very delicate to calculate from ab-initio theory. The assumption made in the calculations of an atomically sharp interface between Fe and W may also be a limiting factor.

\subsection{Analysis of the dynamical structure factor} \label{sec:SQanalysis}

In neutron scattering, the susceptibility can be written in the following form\cite{JensenMackintosh} 

\begin{equation} \label{eq:susc}
\bar{\bar \chi} (\boldsymbol{q}+\boldsymbol{\tau},\omega)=\frac{1}{2} (1+\textrm{cos} \phi) \bar{\bar \chi}_{Ac} (\boldsymbol{q},\omega) + \frac{1}{2}(1-\textrm{cos} \phi) \bar{\bar \chi}_{Op} (\boldsymbol{q},\omega),
\end{equation}
where $\bar{\bar \chi}_{Ac}$ and $\bar{\bar \chi}_{Op}$ are the susceptibilites originating from accoustic and optical branch, respectively, $\boldsymbol{q}$ is a reciprocal vector within the primitive Brillouin zone (BZ), $\boldsymbol{\tau}=[hkl]=h\boldsymbol{b}_1+k\boldsymbol{b}_2+l\boldsymbol{b}_3$, $\phi = \boldsymbol{\tau} \cdot \boldsymbol{\rho}$, where $\boldsymbol{\rho}$ is a vector connecting two sublattices. 
In this manner, by changing the momentum transfer by varying $\boldsymbol{\tau}$, the intensity of the accoustic and optical branches is changing. If we take the Fe bilayer on W(110) as an example, $\boldsymbol{\tau}=[0~ \frac{1}{\sqrt{2}}~  0.5]a$ and the reciprocal vectors are restricted in the film-plane ($l$=0), then it follows that $\boldsymbol{\phi}=h \pi + k \pi$. Inside the primitive BZ, the phase $\phi =0$ and in the limit $\boldsymbol{q} \rightarrow \boldsymbol{0}$, in Eq.~(\ref{eq:susc}) the acoustic term will dominate and will be detected in experiment. If we go outside the first BZ, it is possible to have a situation where the optical term dominates, on the expense of the acoustic response, for instance by choosing $\boldsymbol{\tau}=[1 0]$, as illustrated in Fig.~\ref{fig7b}. If there are more than two atoms in the unit cell, the analysis becomes more complicated but the principle is the same.

\section{Conclusions}

Summarizing, we have shown that combining first principles calculations with atomistic spin dynamics simulations provides a powerful tool for studies of magnetic excitations in low dimensional systems. 
It is hoped that the remarkable progress using the SPEELS technique~\cite{Prokop2009} will continue to provide new and surprising experimental results, for which the currently presented theory seems to be a good tool for analyzing the experimental data and predicting magnon spectra and related properties. The materials presented here have several common features, e.g. the absence of optical modes in the magnon curves, both as determined by experiments as well as obtained by theory. This fact has been analyzed in detail and it is argued that also the optical modes should be visible if one considers excitations which allow for momentum transfer outside the first Brillouin zone. The realization of this in the SPEELS method is clearly a challenge. In addition, we show that all thin film systems investigated here have a spin wave stiffness which is considerably softer compared to the bulk value. This applies both to Fe as well as Co films. Finally, we report on a quantitative agreement between theory and measured magnon curves for 2 ML Fe on W(110) and 8 ML Co on Cu(001).

\appendix
\section*{Appendix}
\label{app}
\setcounter{figure}{0} \renewcommand{\thefigure}{A.\arabic{figure}}

In addition to the \textit{ab initio} calculated values for the magnetic moments and anisotropy energies included in Section~\ref{abinitio}, we present here in detail the values of the exchange interaction divided into intra- and inter-layer contributions.
\begin{figure}[htbp]
                 \includegraphics[scale=0.60]{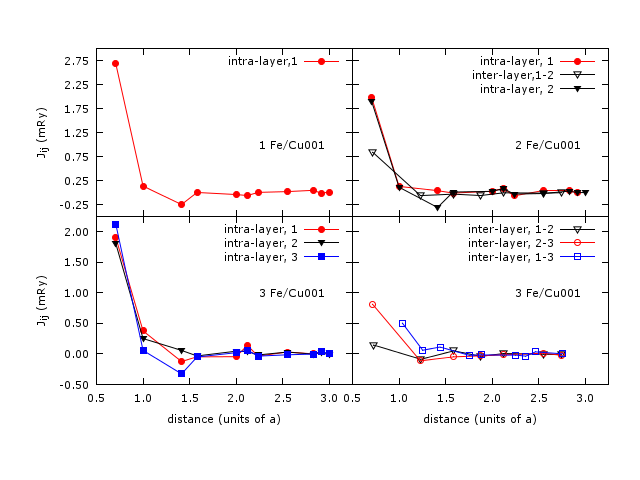}
       \caption{Exchange parameters dependence on the distance (in units of the lattice parameter a). Both the intra- and inter-layer interactions are specified for 1, 2 and 3 layers of Fe on a Cu(001) substrate. The labeling of the Fe layers starts from the interface with the Cu substrate with layer '1' and continues towards the vacuum.}
\end{figure}     
\begin{figure}[tbp]
                 \includegraphics[scale=0.40]{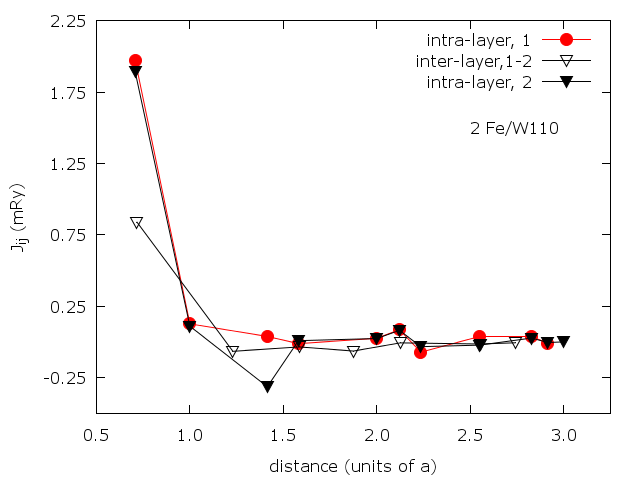}
       \caption{Exchange parameters dependence on the distance (in units of the lattice parameter a). Both the intra- and inter-layer interactions are specified for 2 layers of Fe on a W(110) substrate. The labeling of the Fe layers starts from the interface with the W substrate with layer '1', layer '2' being at the interface with the vacuum.}
\end{figure}     
\vfill
\begin{figure}[ht]
                 \includegraphics[scale=0.60]{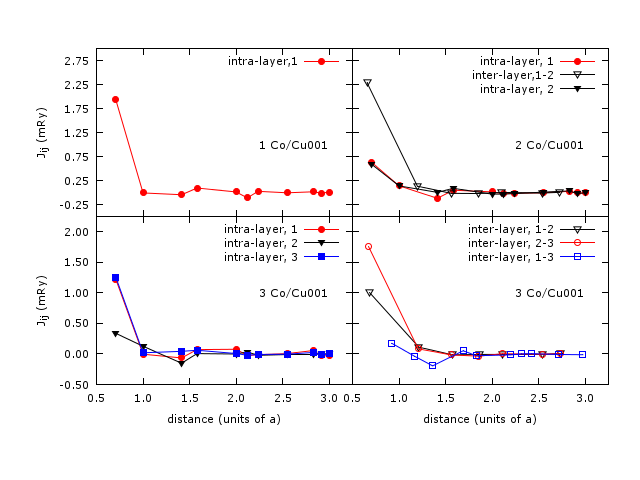}
       \caption{Exchange parameters dependence on the distance (in units of the lattice parameter a). Both the intra- and inter-layer interactions are specified for 1, 2 and 3 layers of Co on a Cu(001) substrate. The labeling of the Co layers starts from the interface with the Cu substrate with layer '1' and continues towards the vacuum.}
\end{figure}     
\begin{figure}[hb]
\centering
                 \includegraphics[scale=0.60]{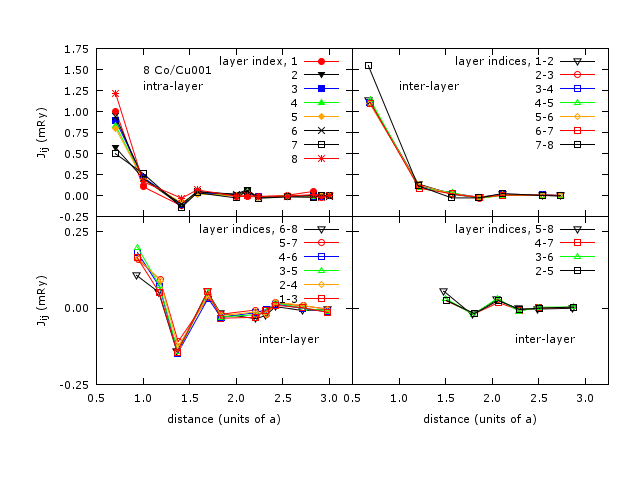}
       \caption{Exchange parameters dependence on the distance (in units of the lattice parameter a). Both the intra- and inter-layer interactions are specified for 8 layers of Co on a Cu(001) substrate. The labeling of the Co layers starts from the interface with the Cu substrate with layer '1' and continues towards the vacuum, layer '8' being the last Co ML.}
\end{figure}     
\vfill  
\clearpage 

\begin{acknowledgments}
We gratefully acknowledge the European Research Council (ERC project 247062 - ASD), the Swedish Research Council (VR), Knut and Alice Wallenberg Foundation, Carl Tryggers Foundation and G\"oran Gustafsson Foundation for financial support.~L.B. and A.B. also acknowledge SeRC and eSSENCE, respectively. The computer simulations were performed on resources provided by the Swedish National Infrastructure for Computing (SNIC) at National Supercomputer Centre (NSC), UppMAX and High Performance Computing Center North (HPC2N). We thank B. Hj\"orvarsson, J. Kirschner, L. Nordstr\"om, E. Papaioannou, L. Szunyogh, L. Udvardi, D. Mills and Kh. Zakeri for valuable discussions.
\end{acknowledgments}

\end{document}